\documentclass[useAMS,usenatbib,letterpaper]{mn2e}

\usepackage{amsmath,mathtools,graphicx,amssymb,verbatim,url,epsfig,natbib,bm,psfrag,ifthen,hyperref}
\usepackage[english]{babel}
\usepackage{epsfig}
\usepackage{times}
\usepackage{pifont}
\usepackage[nomessages]{fp}

\renewcommand{\vec}[1]{\mathbf{#1}}

\DeclareMathOperator*{\size}{\ensuremath{R_{g \ast p}/R_{p}}}
 
\DeclareMathOperator*{\tr}{^\intercal}
\providecommand{\mean}[1]{\ensuremath{\langle #1 \rangle}}
\providecommand{\std}[1]{\ensuremath{std(#1)}}

\DeclareMathOperator*{\argmin}{arg\,min}

\title[S\'{e}rsic galaxy models in weak lensing shape measurement]
{
S\'{e}rsic galaxy models in weak lensing shape measurement: model bias, noise bias and their interaction
}

\author[Tomasz Kacprzak et al.]{
Tomasz~Kacprzak,$^{1}$\thanks{E-mail: \texttt{tomasz.kacprzak.09@ucl.ac.uk}}
Sarah~Bridle,$^{2}$
Barnaby~Rowe,$^{1}$\newauthor
Lisa~Voigt,$^{1}$
Joe~Zuntz,$^{2,3,4}$
Michael~Hirsch$^{1,5}$,
Niall~MacCrann$^{1,2}$
\\
$^{1}$Department of Physics \& Astronomy, University College London, Gower Street, London WC1E 6BT\\
$^{2}$Jodrell Bank Centre for Astrophysics, University of Manchester, Manchester, M13 9PL \\
$^{3}$Astrophysics Group, University of Oxford, Denys Wilkinson Building, Keble Road, Oxford OX1 3RH\\
$^{4}$Oxford Martin School, University of Oxford, Old Indian Institute, 34 Broad Street, Oxford OX1 3BD\\
$^{5}$Max Planck Institute for Intelligent Systems, Department of Empirical Inference, Spemannstra\ss{}e 38, 72076 T\"ubingen, Germany \\
}

\begin{document}

\date{\today}

\pagerange{\pageref{firstpage}--\pageref{lastpage}} \pubyear{2011}

\maketitle

\label{firstpage}

\begin{abstract}
Cosmic shear is a powerful probe of cosmological parameters, but its potential can be fully utilised only if galaxy shapes are measured with great accuracy.
Two major effects have been identified which are likely to account for most of the bias for maximum likelihood methods in recent shear measurement challenges. 
Model bias occurs when the true galaxy shape is not well represented by the fitted model.
Noise bias occurs due to the non-linear relationship between image pixels and galaxy shape.
In this paper we investigate the potential interplay between these two effects when an imperfect model is used in the presence of high noise.
We present analytical expressions for this bias, which depends on the residual difference between the model and real data.
They can lead to biases not accounted for in previous calibration schemes.

\noindent
By measuring the model bias, noise bias and their interaction, we provide a complete statistical framework for measuring galaxy shapes with model fitting methods from GRavitational lEnsing Accuracy Testing (GREAT) like images.
We demonstrate the noise and model interaction bias using a simple toy model, which indicates that this effect can potentially be significant.
Using real galaxy images from the Cosmological Evolution Survey (COSMOS) we quantify the strength of the model bias, noise bias and their interaction.
We find that the interaction term is often a similar size to the model bias term, and is smaller than the requirements of the current and shortly upcoming galaxy surveys.
\end{abstract}


\section{Introduction}
\label{sec:introduction} 

Weak gravitational lensing is a very important and promising probe of cosmology \citep[see][for reviews]{Schneider1996,BartelmannSchneider2001,HoekstraJain2008}. 
Matter between a distant galaxy and an observer causes the image of the galaxy to be distorted.
This distortion is called gravitational lensing.
Almost all distant galaxies we observe are lensed, mostly only very slightly, so that the observed image is sheared by just few percent. 
Weak gravitational lensing shear is particularly effective in constraining cosmological model parameters \citep{detfr2006,esoesa,Fu+2008,Kilbinger+2013}.
Measuring the spatial correlations of those shear maps in the tomographic bins of redshift can shed light on the evolution of dark energy in time \citep{hu2002,TakadaJain2004,huff+2011b,heymans+2013,Benjamin+2013} and modified gravity \citep{Simpson+2013,Kirk+2013}.

Several projects are planning to measure cosmic shear using optical imaging.
The KIlo-Degree Survey (KIDS), 
the Dark Energy Survey (DES)\footnote{http://www.darkenergysurvey.org},
the Hyper Suprime-Cam (HSC) survey\footnote{http://www.naoj.org/Projects/HSC/HSCProject.html},
the Large Synoptic Survey Telescope (LSST)\footnote{http://www.lsst.org}, 
Euclid\footnote{http://sci.esa.int/euclid} 
and Wide Field Infrared Survey Telescope (WFIRST)\footnote{http://exep.jpl.nasa.gov/programElements/wfirst/}. 

However, accurate measurement of cosmic shear has proved to be a challenging task \citep{step1,step2,great08results,great10results}.
There is a range of systematic effects that can mimic a shear signal.
In this paper, we focus on biases in the measurement of shear.
Other important systematics are: intrinsic alignments of galaxy ellipticities, photometric redshift estimates and modelling of the clustering of matter on  small scales in the presence of baryons.

Prior to shearing by large scale structure, galaxies are already intrinsically elliptical. 
This ellipticity is oriented randomly on the sky in the absence of intrinsic alignments. 
Then a few percent change in this ellipticity is induced as the light travels from the galaxy to the observer through intervening matter. 
During the observation process, the images are further distorted by a telescope Point Spread Function (PSF) and, in case of ground-based observations, also by atmosphere turbulence. 
Additionally the image is pixelised by the detectors.
Due to the finite number of photons arriving on the detector during an exposure, the galaxy images are noisy.
Additional noise is induced by the CCD readout process in the detector hardware.

The complexity of this forward process makes the unbiased measurement of the shear signal very challenging.
\citet{BonnetMellier1995} showed that for a very good quality data, simple quadrupole moments of the images can be used as unbiased shear estimators. 
Example methods utilising this approach are \citet{ksb1,Kaiser2000,HirataSeljak2003,Okura+2011}.
For PSF convolved galaxy images, one can use DEconvolution In MOment Space \citep[DEIMOS][]{deimos1} to remove the effects of the PSF from the quadrupole. 

Model fitting methods use a parametric model for the galaxy image to create a likelihood function (often multiplied by the prior), and then extract a ellipticity from it. Galaxy images are often modelled by S\'{e}rsic functions:
\textsc{Im3shape} \citep{im3shape} uses a 7-parameter bulge + disc model and and infers the parameter values by maximum likelihood estimation.
\citet{lensfit1,lensfit3} uses a similar model, and mean posterior for the estimator.
Another frequently used galaxy model is a decomposition into a Gauss-Laguerre orthogonal set \citep{shapelets1,BernsteinJarvis2002,GL1}, `shapelets'. The complexity of this model can be controlled by changing the number of coefficients in the shapelet expansion.

In the context of model fitting methods, if the model is not able to represent realistic galaxy morphologies well enough, then the shape estimator will be biased \citep{bernstein2010,Voigt2011}. This bias is often called the \emph{model bias} or \emph{underfitting bias}.

In the presence of pixel noise on the image, the unweighted quadrupole moment will give an unbiased estimate of the quadrupole, with a very large variance. 
However, the shear is defined as a ratio of quadrupole moments, and \citet{hirataetal04,melchior2012,Okura+2013} showed that this induced non-linearity leads to a bias. This bias is often called the \emph{noise bias}.

For model fitting methods, the noise related bias was studied in \citet[][hereafter R12]{NoiseBias1}, where analytical expressions were given for the estimator bias.
It presented the noise bias as a simple statistical problem, assuming that the true galaxy model was perfectly known.
The noise was added to images which were created by the same model function which then later was used to fit it.
In \citep[hereafter K12]{nbc_paper} we have used the S\'{e}rsic galaxy models to quantify the magnitude of noise bias and presented a calibration scheme to correct for it. 
Again, here the true galaxy model was perfectly known.
In \citet{im3shape} we have further developed the calibration scheme. 
It was then applied to the GREAT08 simulation set \citep{great08handbook,great08results}, with satisfactory results on most challenge branches.

In this paper we investigate the next piece in the puzzle: what if model biases and noise biases occur at the same time?
This will certainly be the case for real galaxy surveys: the galaxies have realistic morphology and the images are noisy. 
In that situation there may be some interaction between noise and model bias. 

\begin{figure*}
\label{fig:concept}
\epsfig{file=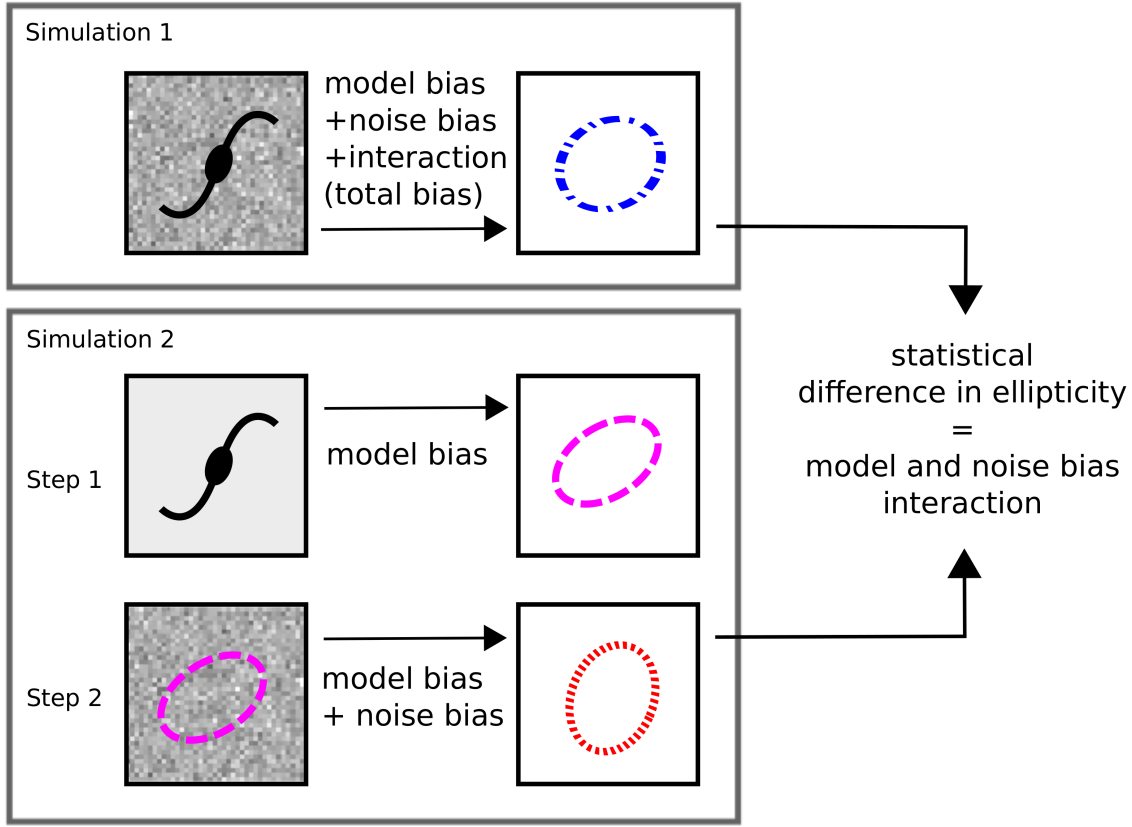,width=16cm}
\caption{A demonstration of the concept of noise and model bias interaction. 
Left-hand side postage stamps show the images which are going to be fit with a parametric galaxy model.
Right-hand side postage stamps are the images of best fitting models.
In this work we compare the results of two simulations.
Simulation 1 uses the real galaxy image directly to measure noise bias, model bias and their interaction jointly in a single step (blue dot-dashed ellipse).
Simulation 2 first finds the best fitting parametric model to a real galaxy image (black solid galaxy cartoon), and create it's model image (dashed magenta ellipse).
This process introduces the model bias.
The next step is to measure the noise bias using this best fitting image as the true image.
The fit to the noisy image is represented by red dotted ellipse.
In this simulation noise and model bias interaction terms are absent.
The difference of the results of Simulations 1 and 2 measures the strength of noise and model bias interaction terms.
}
\end{figure*}

Figure \ref{fig:concept} demonstrates this concept. Is shows two effects studied so far:
\begin{itemize}
	\item \emph{model bias} - when the galaxy image is noiseless, and the fitted model does not represent the complicated galaxy features well 
	\item \emph{noise bias} - when galaxy image is noisy, and the fitted model is perfectly representing the galaxy
\end{itemize}
and the effect we investigate more in this work:
\begin{itemize}
	\item \emph{model bias, noise bias and their interaction} 
- when the true galaxy has realistic morphology, the fitted model does not represent the galaxy well, and the observed image is noisy.
\end{itemize}

We present analytic equations for the noise bias when the true galaxy model is not perfectly known.
Then we use a toy model to show that the interaction terms have the potential to be significant.
To evaluate the significance of this interaction terms we use 26113 real galaxy images from the COSMOS survey \citep{shera}, available in the \textsc{GalSim}\footnote{\url{github.com/GalSim-developers/GalSim}} toolkit (Rowe et al. 2013, in prep.). 

Therefore, by evaluating the model bias, noise bias and their interaction, we attempt to answer the question: 
\emph{can we use S\'{e}rsic profiles to represent galaxies in fitting to realistic noisy data?}
However, our answer is limited to a simple bulge + disc galaxy model, which is commonly used in model fitting shape measurement methods for weak lensing. 
Also, we use the PSF and pixel size that corresponds to the parameters of 
a conservative upcoming ground-based Stage III survey.
It is possible to perform similar analysis for different galaxy models and telescope parameters, as well as for different types of estimator (maximum posterior, mean posterior).

Calibration of biases in shear measurement is becoming an increasingly important part of shear pipelines. 
\citet{lensfit3} used a calibration for noise bias as a function of galaxy size and signal to noise ratio.
Systematic effects may depend on many different galaxy and image quality properties, and increasing the accuracy of the bias measurement will require including more of these effects into the pipeline. 
This will increase the complexity of the modelling and 
the computational cost of the simulation.
Moreover, it will require knowledge of the true underlying parameters for a galaxy sample, either in the form of a representative calibration sample of images, or an inferred galaxy parameters distribution.
\citet{ControlLoops} introduced a procedure to infer these parameters via a Monte Carlo Control Loops approach.
So far, in various simulations, galaxy images were often created using a parametric S\'{e}rsic model function.
An important question is: are the realistic galaxy morphologies important for shear bias measurement? The upcoming GREAT3 challenge (Mandelbaum et al. 2013, in prep) is planning to answer this question by testing using images of galaxies in the COSMOS survey. 
In this paper we provide an initial answer to this question. 
A more detailed study of galaxy model selection is going to be presented in (Voigt et al. 2013, in prep.) and calibration sample requirements in (Hirsch et al. 2013 in prep.).

It is worth to note that some shear measurement methods apply a fully Bayesian formalism and use a full posterior shear probability \citep{BernsteinArmstrong2013} in subsequent analyses.
These methods are considered to be free of noise and model related biases and present a promising alternative approach to problems studied in this paper.

This paper is organised as follows. 
Section \ref{sec:basics} contains the principles of cosmic shear analysis. 
In Section \ref{sec:analytic} we present the analytic formulae for noise and model bias interaction, as a generalisation of the noise bias equations derived in R12.
We present a toy model for the problem in section \ref{sec:toy}. 
In Section \ref{sec:eval} we use the COSMOS sample to evaluate the noise and model biases, and show the significance of the interaction terms.
We conclude 
in Section \ref{sec:summary}.

\section{Systematic errors in model fitting} 
\label{sec:basics}
In this section, first we present the basics of the model fitting approach to shear measurement and then discuss the biases it introduces. We discuss the requirements on this bias in the context of current and future surveys.
Then we introduce analytical expressions for the noise bias, including interaction terms with the model bias.

\subsection{Shear and ellipticity} 
Cosmic shear as cosmological observable can be related to the gravitational potential between the distant source galaxy and an observer \citep[see][for reviews]{BernsteinJarvis2002}.
Ellipticity is defined as a complex number
\begin{equation}
e=\frac{a-b}{a+b}\,\,e^{2i\phi}, 
\label{eqn:ellipticity_definition}
\end{equation}
where $a$ and $b$ are the semi-major and semi-minor axes, respectively, and $\phi$ is the angle (measured anticlockwise) between the x-axis
and the major axis of the ellipse. The observed (lensed) galaxy ellipticity is modified by the complex shear $g=g_{1}+ig_{2}$ in the following way
\begin{equation}
e^{l}=\frac{e^i+g}{1+g^{*}e^{i}}.
\label{eqn:eobs}
\end{equation}
In the absence of intrinsic alignments the lensed galaxy ellipticity is an unbiased shear estimator \citep{seitzs97}.

To recover this ellipticity, the shear measurement methods correct for the PSF and the pixel noise effects.
This procedure can introduce a bias. 
Usually shear bias is parametrised with a multiplicative $m$ and additive $c$ component
\begin{equation}
\label{eqn:shear_bias_def}
\hat{\gamma}_j=(1+m_j) \gamma^{t}_j+c_j, 
\end{equation}
where $\hat{\gamma}$ is the estimated shear and $\gamma^{t}_j$ is the true shear. 
Additive shear bias is usually highly dependent on the PSF ellipticity and can be probed using the star-galaxy correlation function \citep{lensfit3}, which can be calculated on the measurements from the real data.
Efforts are made to calibrate it using simulations \citep[K12]{lensfit3}.
Requirements for multiplicative and additive bias are summarised in Table \ref{tab:requirements}, derived using \citet{amara2008}.

\begin{table}
\center
\begin{tabular}{|c|ccc|ccc|ccc|}
\hline
Survey   		& area (sq deg) & $m_i$ & $c_i$  \\
\hline
Current  		& 200          & 0.02  & 0.001   \\
Upcoming future & 5000		   & 0.004 & 0.0006  \\ 
Far future   	& 20000		   & 0.001 & 0.0003  \\ 
\hline
\end{tabular}
\caption{Requirements for the multiplicative and additive bias on the shear for current, upcoming and far future surveys, based on \citet{amara2008}
\label{tab:requirements}
}.

\end{table}

When fitting a galaxy model with co-elliptical isophotes, this ellipticity is often amongst the model parameters.
An estimator is created using a likelihood function, which in the presence of white Gaussian noise has the form
\begin{align}
-2 \log \mathcal{L} &= \frac{1}{\sigma^2} \sum_p \left[ g_p + n_p - f_p(\vec{a}) \right]^2  
\label{eqn:likelihood} 
\end{align}
where $p$ is the pixel index running from $1$ to $N$, where $N$ is the number of pixels, 
$\vec{a}$ is a set of variable model parameters, 
$g_p$ is the noiseless galaxy image, 
$n_p$ is additive noise with standard deviation $\sigma_n$ and
$f_p$ is a model function.

In this approach, the two main systematic effects are \emph{noise bias} and \emph{model bias}.
Model bias has so far been
studied in the context of low noise (or noiseless) images. 
\citet{lewis2009,Voigt2011,bernstein2010} demonstrated that fitting galaxy image with a model which consists of a basis set (or degrees of freedom) smaller than in the true galaxy image, then the model fitting method can be biased. 
This bias can even be larger than the current survey requirements.
Voigt et al. (2013, in prep) quantify the model bias using real galaxy images from the COSMOS survey, for different galaxy models used in the fit.

Noise bias arises when the observed galaxy images contain pixel noise.
K12 showed that for galaxy with noise level of $S/N>200$ noise bias is negligible.
However, for galaxies with $S/N \approx 20$ it can introduce a multiplicative bias up to $m=0.08$, 
which is large compared to our requirements for future surveys (see Table \ref{tab:requirements}). 
This is a very important systematic effect that will have to be accounted for to utilise the full statistical power of a survey.
 
R12 showed analytic expressions for the noise bias in the case when the true model is perfectly known. 
The key idea was that the value of the parameter estimator obtained from the noisy data lies near the true value for this parameter, so an expansion can be used to calculate it. 
It turns out that the terms which depend on the first power of noise variance disappear when averaged over pixel noise realisations.
Second order terms, however, give rise to the noise bias.
The bias was quantified for a simple Gaussian galaxy model using both analytical expressions and simulation, which were found to be in good agreement.

In the following section we expand the derivation in R12 to the case when the true galaxy model is not known. 

\section{Analytic expressions for noise and model bias, with their interaction} 

We introduce a generalisation of noise bias equations in R12 to include the case when the galaxies measured have unknown morphologies. 
We use an expansion around the \emph{parameters of the best fit model to the noiseless data}.
Note that in this formalism there is no \emph{true} value of the parameters, as the real galaxy can have morphology which is not captured by the model. 
This in fact gives rise to model bias, which has to be evaluated empirically from low noise calibration data.

In Appendix \ref{sec:appendix1} we derive the noise bias when an incomplete model is used.
This derivation is summarised here. 

Let us use the galaxy model function $f(\vec{a})_p$ for pixel $p$, and Fisher matrix $F$, Jacobian $D^{(1)}_{ip}$, a Hessian for each pixel $D^{(2)}_{ijp}$, and third derivative tensor $D^{(3)}_{ijkp}$
\label{sec:analytic}
\begin{align}
\label{eqn:D_def}
D^{(1)}_{ip} &:= \frac{\partial f(\vec{a}^t)_p }{ \partial a_i } \\ 
D^{(2)}_{ijp} &:= \frac{\partial^2 f(\vec{a}^t)_p }{ \partial a_i \partial a_j } \\
D^{(3)}_{ijkp} &:= \frac{\partial^3 f(\vec{a}^t)_p }{ \partial a_i \partial a_j \partial a_k} \\
F_{ij} &:= \frac{\partial f(\vec{a}^t)_p }{ \partial a_i } \frac{\partial f(\vec{a}^t)_p }{ \partial a_j } =  D^{(1)}_{ip} D^{(1)}_{jp}
\end{align}
where $\vec{a}^t$ is a parameter vector which produces a best fit to the noiseless image and summation over repeated indices is used.

The covariance between two parameters of our model is
\begin{align}
\langle a^{(1)}_i a^{(1)}_j \rangle =  \sigma^2 \tilde F_{ij}^{-1} F_{ij} \tilde F_{ij}^{-1}
\end{align}
where $\tilde F_{ik} := (F_{ik}  - r_p D^{(2)}_{ikp})$ is a Fisher matrix modified by the residual of the best fit to the noiseless real galaxy image and our model 
\begin{align}
\label{eqn:residual}
r_p:=g_p - f_p(\vec{a}^t).
\end{align}
Note that for no model bias $r_p=0$ and $\langle a^{(1)}_i a^{(1)}_j \rangle = F_{ij}$.

The bias on the parameter estimate is 
\begin{align}
\langle a^{(2)}_i \rangle 
&=
  \sigma_n^2  \tilde F_{ik}^{-1} \big[ \nonumber \\
+ &\tilde F^{-1}_{lj} D^{(1)}_{jp} D^{(2)}_{lkp} \nonumber \\
+ \frac{1}{2} & \tilde F^{-1}_{lj} F_{lj} \tilde F^{-1}_{lj}  D^{(3)}_{ljkp} r_p  \nonumber \\
- &\tilde F^{-1}_{lj} F_{lj} \tilde F^{-1}_{lj} D^{(1)}_{jp} D^{(2)}_{lkp} \nonumber \\
- \frac{1}{2} &\tilde F_{lj}^{-1} F_{lj} \tilde F^{-1}_{lj} D^{(1)}_{kp} D^{(2)}_{ljp} 
 \big] 
\label{eqn:nmb_final}
\end{align}
Again, if $r_p = 0$, then the expression reduces to the R12 result:
\begin{align}
\label{eqn:nb_final}
\langle a^{(2)}_i \rangle 
&=
\sigma_n^2  F_{ik}^{-1} \big[ \nonumber \\ 
&+ F^{-1}_{lj} D^{(1)}_{jp} D^{(2)}_{klp}  \nonumber \\
&- F^{-1}_{lj} D^{(1)}_{jp} D^{(2)}_{klp}  \nonumber \\
&- \frac{1}{2} F^{-1}_{lj} D^{(1)}_{kp} D^{(2)}_{ljp}  
\big] \nonumber \\
&= - \frac{1}{2} \sigma_n^2  F_{ik}^{-1} F^{-1}_{lj} D^{(1)}_{kp} D^{(2)}_{ljp}  .
\end{align}

These expressions indicate that the residual image modifies the expressions for the noise bias.
It prevents a cancellation of two terms and introduces an additional term in $D^{(3)}_{ljkp}$.

\section{Toy model for the problem} 
\label{sec:toy}

\begin{figure*}
\epsfig{file=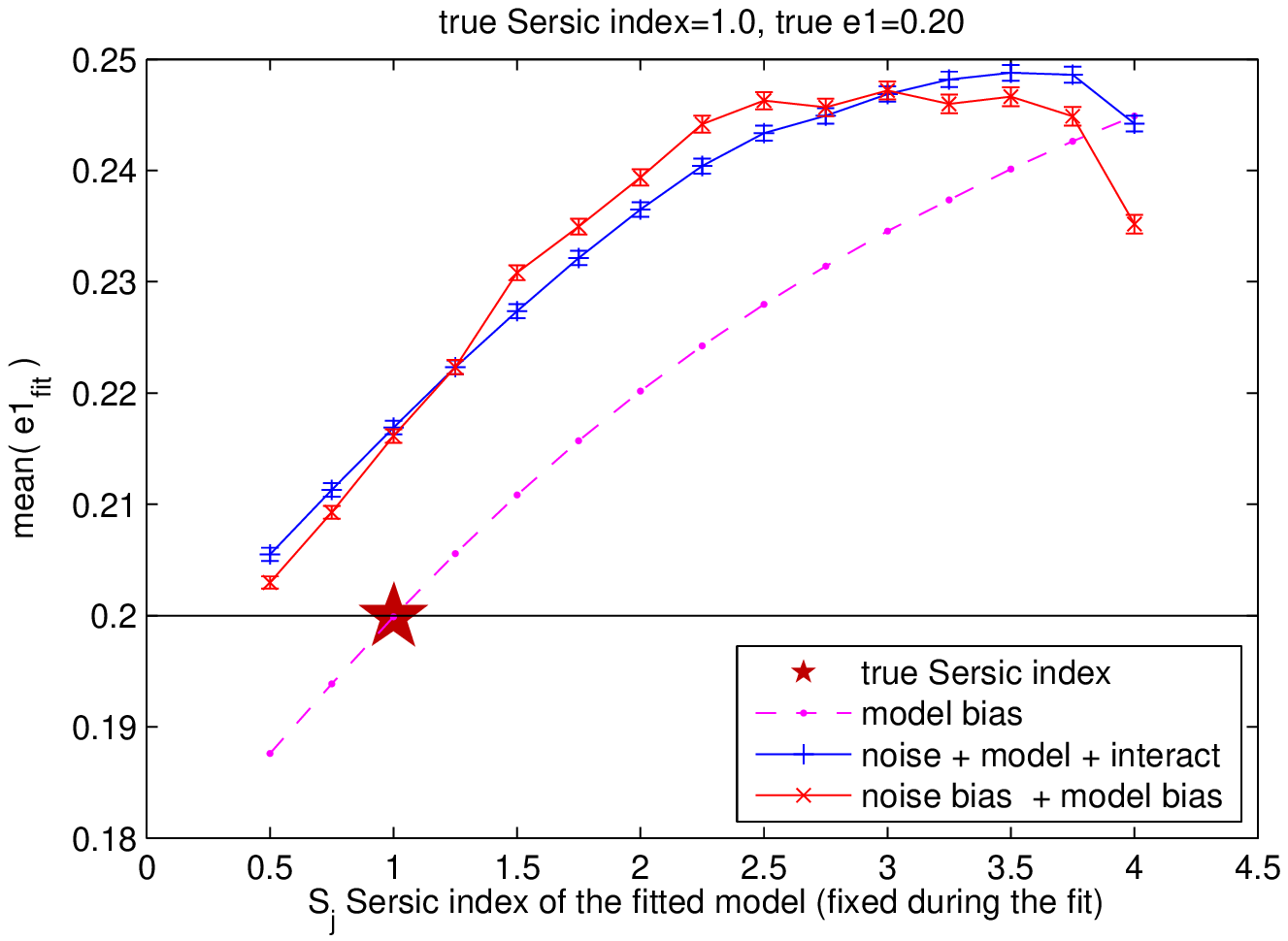,width=10cm}
\epsfig{file=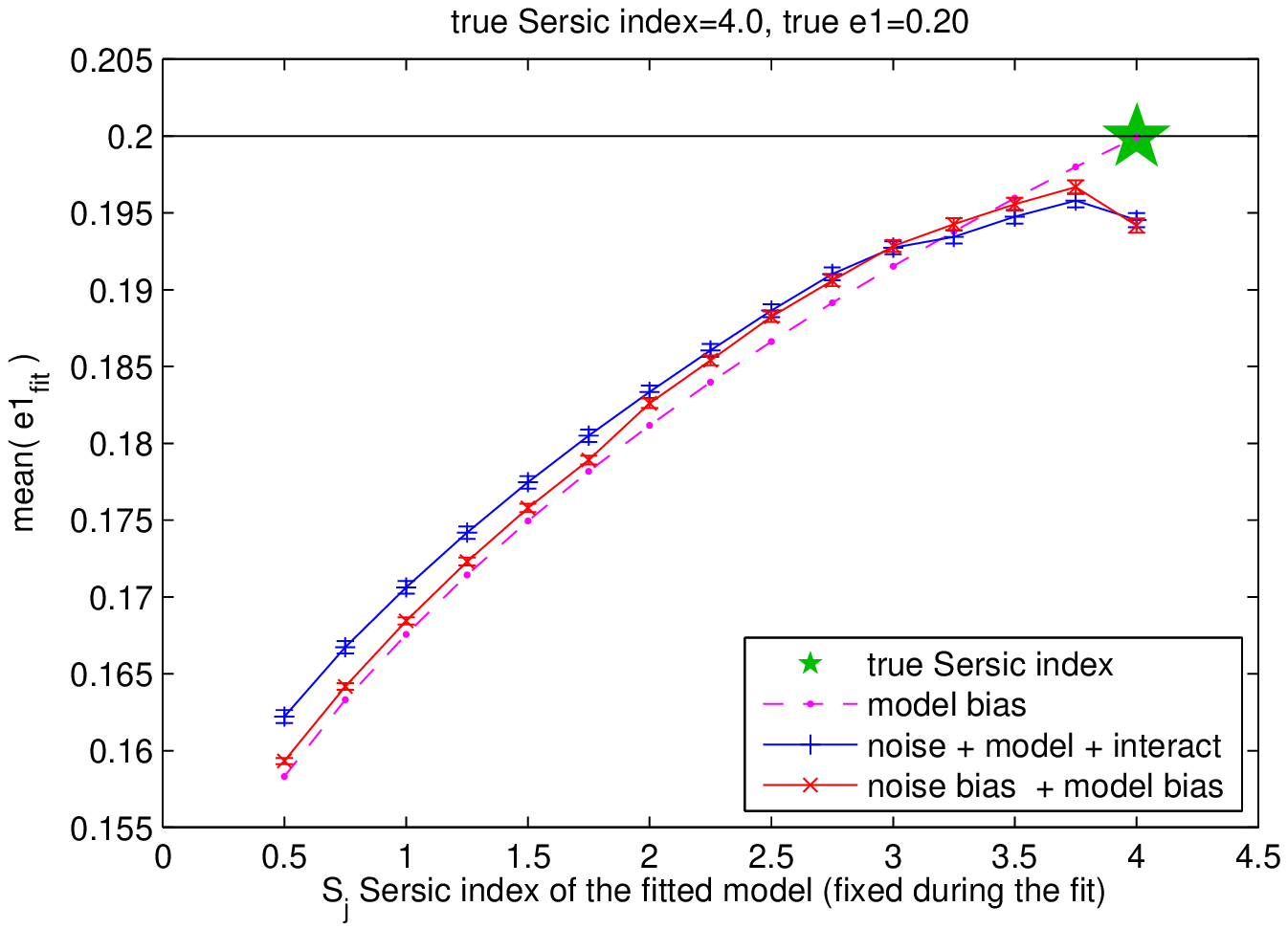,width=10cm}
\epsfig{file=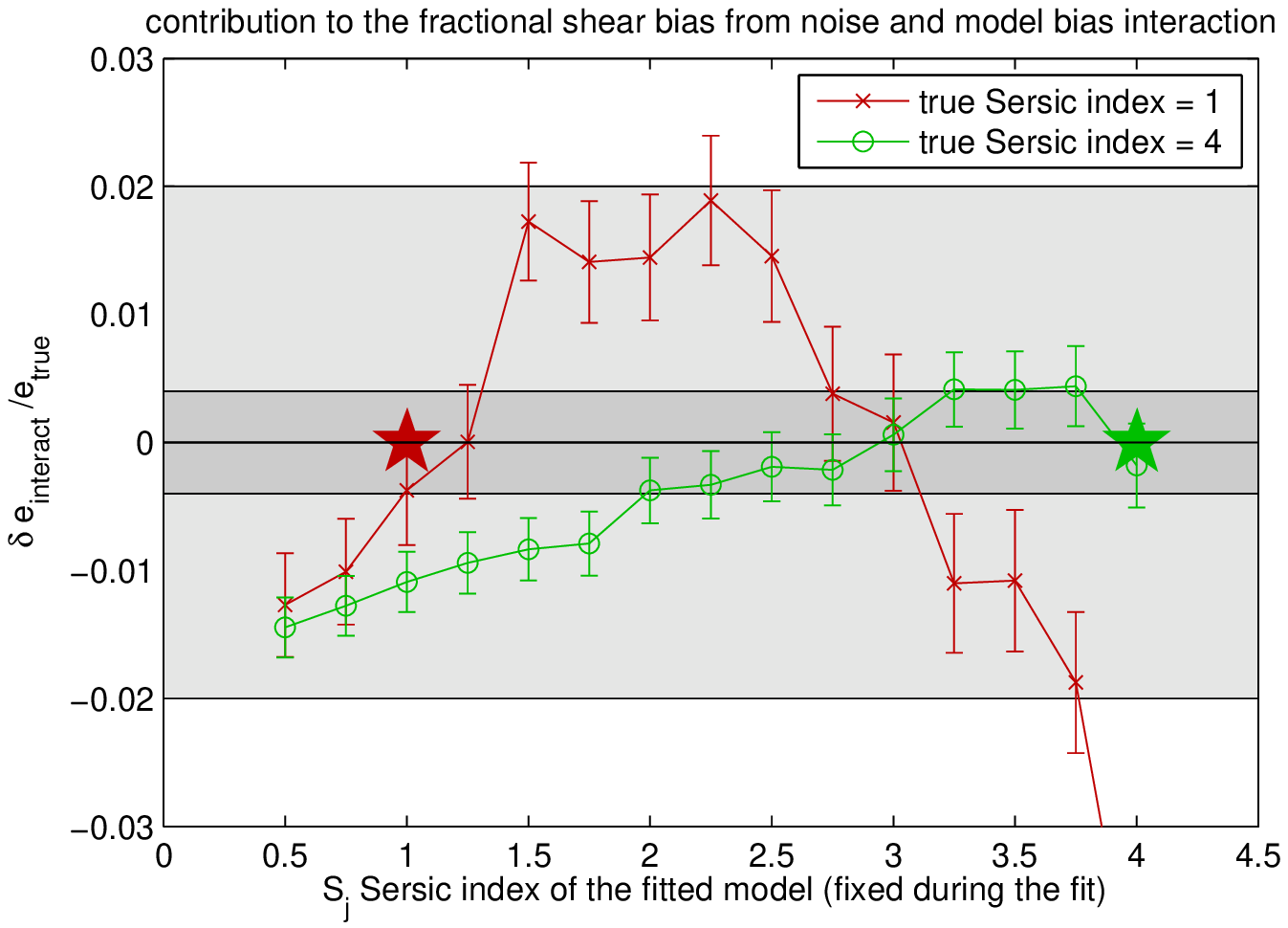,width=10cm}
\caption{Noise and model bias toy model.
The upper and middle panels show the galaxy ellipticity estimated when a wrong S\'{e}rsic index $S_j$ is used.
True ellipticity was $e_1^{true}=0.2$. 
True S\'{e}rsic index was 1 and 4 for the top and middle panels respectively.
Additionally, a star marks the true S\'{e}rsic index, to guide the eye.
Magenta dash-dotted line with cross dot markers shows the model bias; ellipticity estimated by a fit with S\'{e}rsic $S_j$ in the absence of noise.
Blue and red lines with $+$ and $\times$ markers, respectively, show the mean ellipticity measured from galaxy image with added noise, when:
(1) galaxy image has the true S\'{e}rsic index, 
(2) galaxy image is the best fit with S\'{e}rsic index $S_j$.
Bottom panel shows the fractional difference between $[(1)-(2)] / e_1^{true}$.
Grey bands mark the requirement on multiplicative bias for current and upcoming surveys.
\label{fig:toy_model}
}
\end{figure*}

To demonstrate the noise and model bias interaction we create a simple toy example.
Galaxy models used in this example will be single S\'{e}rsic profiles \citep{sersic63}, where surface brightness at image position $\vec{x}$:
\begin{equation}
I(\vec{x})= A\exp( -k [(\vec{x}-\vec{x_0})^{\tr} \vec{C}^{-1} (\vec{x}-\vec{x_0})] ^{1/(2n)} )
\end{equation}
where $\vec{x_o}$ is galaxy centeriod, $\vec{C}$ is galaxy covariance matrix (see e.g. \citet{voigt2010} for relation to ellipticity), $k=1.9992n-0.3271$ and $n$ is the S\'{e}rsic index.

Instead of a real galaxy, we use a single S\'{e}rsic profile with a fixed index.
Hereafter we will refer to that galaxy as the \emph{real} galaxy, as it will serve as a real galaxy equivalent in our toy example, even though it is created using a S\'{e}rsic function.
We fit it with another single S\'{e}rsic profile, but with a different index. 
In this section we will refer to it as the \emph{galaxy model}.
This way we introduce model bias - the real galaxy is created using a S\'{e}rsic profile with a different index than the model we fit.

The impact of the noise and model interaction bias terms introduced in Eqn. \ref{eqn:nmb_final} can be measured in the following way (Fig. \ref{fig:concept}).
We use the real galaxy, add noise to it, and measure the ellipticity by fitting the galaxy model.
This process will include the noise bias, model bias and their interaction terms, which are dependent on the pixel residual between the best fit to the noiseless image and the real galaxy image (Eqn. \ref{eqn:residual}). This process is described in Simulation 1 in Fig \ref{fig:concept}.
These residuals will be zero if instead of the real galaxy image we use the best-fitting galaxy model (Simulation 2 in Fig \ref{fig:concept}). 
This is the pure noise bias scenario, as described in Eqn. \ref{eqn:nb_final}.
To measure the model bias and noise bias jointly, we use a two step procedure. 
In step 1 we obtain the best fit image to the noise-free real galaxy.
Thus we obtain the model bias measurement.
In step 2 we use this model image to measure the mean of noise realisations by repeatedly adding noise to it.
The strength of the noise and model bias interaction terms scaled by the residual $r_p$ in Eqn. \ref{eqn:nb_final} can be measured by taking the difference between the mean of the noise realisations from the cases when the true image is a \emph{real} galaxy and when the true image is the \emph{best-fitting} galaxy, obtained earlier.

Fig. \ref{fig:toy_model}, upper and middle panels 2 show these cases when the real galaxy is a S\'{e}rsic with an index of 1 and 4, respectively. 
We used a Moffat PSF with $\text{FWHM}=2.85$ pixels, and kept the half light radius of these galaxies so that the FWHM of the convolved object divided by the FWHM of the PSF is 1.4. 
These settings were used in the GREAT08 challenge and in previous noise bias work.
The true ellipticity of the real galaxy was set to $e_1^{\mathrm{true}} = 0.2$.
For the noisy cases, we used $S/N=20$, where ${\rm S/N} = \sqrt{\sum_{p=1}^N g_p^2} / {\sigma_\mathrm{noise}}$ \citep[see][]{great08results}.

Fig. \ref{fig:toy_model} shows the mean measured ellipticity $e_1^{\mathrm{fit}}$ as a function of the S\'{e}rsic index of the fitted model. 
The magenta dashed line shows the pure model bias measurement.
There was no noise on the images in this case: the mean is taken from the average of 
simulated
images, which had a centroid randomly sampled within a central pixel in the image. 
Standard errors for this measurement are so small that they are not visible on this plot.
We see that if the correct model is used, then the measured ellipticity is equal to the true ellipticity $e_1^{\mathrm{true}}=0.2$. 
This point is additionally marked with a star.
For the real galaxy case with a S\'{e}rsic index of 4 (middle panel), ellipticity is underestimated when the index of the fit is smaller than 4. 
The same applies to the case when the S\'{e}rsic index of the real galaxy is 1. 
For that case, ellipticity is overestimated if the fitted index is greater than 1.

The mean of the noisy realisations for the noise bias + model bias + their interaction is marked by a blue solid line with plus $+$ markers.
For this case we were adding noise to the real galaxy images (S\'{e}rsic profiles with indices 1 for the top panel and 4 for the middle panel).
Noise bias effects cause overestimation of ellipticity, relative to the underlying model bias, for fitted S\'{e}rsic indices of $0.5-3$.
For indices $3-4$ the ellipticity becomes underestimated.

The red solid line with cross $\times$ markers shows the model bias and noise bias without the interaction terms.
To obtain it, we use the best fit model image to the noiseless real galaxy with a given S\'{e}rsic index (x-axis).
Then we add noise realisations with the same noise variance as in the case above.
The mean of noise realisations is marked with the red line and cross points.
The interaction terms cause the blue and red lines to differ.
When the true model is used in the fit (S\'{e}rsic index 1 and 4 for the left and right panels respectively), then noise and model interaction bias terms are zero, and mean biases are consistent.

The bottom panel shows the difference between the noise + model + interaction measurement bias and the noise + model only bias, which probes the strength of the interaction terms. 
In this case we plotted this difference as the fractional bias, $\delta e_{\mathrm{interact}}/e_{\mathrm{true}}$.
This fractional bias will be similar to the multiplicative bias in Eqn. \ref{eqn:shear_bias_def} measured from a ring test \citep{GL1}.
Red crosses and green circles correspond to cases where the true S\'{e}rsic index was 1 and 4, respectively.
The current and upcoming survey requirements are shown as light and dark grey bands, respectively.
In the toy example, noise and model interaction bias terms in Eqn. \ref{eqn:nmb_final} can be as significant as $\delta e_{\mathrm{interact}}/e_{\mathrm{true}} = 0.02$.
This demonstrates the potential significant impact of effect: this contribution can exceed the upcoming survey requirements.

The difference is smaller than the effects of the noise bias or model bias alone.
In fractional terms, the noise bias bias on ellipticity was of order $\delta e_{\mathrm{noise}} = (\mean{e_{\mathrm{fit}}} - e_{\mathrm{true}})/e_{\mathrm{true}} = 0.1$.
The maximum model bias measured in this example was $\delta e_{\mathrm{model}} = (\mean{e_{\mathrm{fit}}} - e_{\mathrm{true}})/e_{\mathrm{true}} = 0.2$ when the real galaxy was created using S\'{e}rsic index $S_j=1$ and fitted model galaxy had S\'{e}rsic index of $S_j=4$.

We have demonstrated that using the wrong galaxy model can give rise to significant noise and model interaction bias terms.
However, this bias will strongly depend on the model we use and the real galaxy morphologies.
In next section we quantify the strength of the model bias, noise bias and their interaction using real galaxies from the COSMOS survey, and use a more realistic bulge + disc galaxy model to perform the fit.

\section{Biases for real galaxies in the COSMOS survey}
\label{sec:eval}

\begin{table*}
\begin{tabular}{|c|c|c|c|c|c|c|c|c|}
\hline
true image		      &  $m_1$  &  $m_2$  &  $\std{m_1}$ & $\std{m_2}$ & $c_1$   & $c_2$   & $\std{c_1}$ & $\std{c_2}$ \\
\hline
real galaxies         &  0.0238 &  0.0227 &  0.0006      & 0.0006      & -0.0019 & -0.0018 &  0.0001     &  0.0001 \\
best-fitting galaxies &  0.0230 &  0.0218 &  0.0007      & 0.0006      & -0.0017 & -0.0017 &  0.0001     &  0.0001 \\
difference            &  0.0008 &  0.0009 &  0.0009      & 0.0009      & -0.0001 & -0.0001 &  0.0001     &  0.0001
\end{tabular}
\caption{Multiplicative and additive biases measured from the noise realisations with $S/N=20$. 
The difference is the strength of the model and noise bias interaction terms.
\label{tab:total_bias}
}
\end{table*}

In this section we evaluate the strength of the noise and model bias interaction terms using real galaxy images from the COSMOS survey \citep{shera} available in \textsc{GalSim} (Rowe et al, 2013, in prep.). 
The comparison is done in a very similar way to that in Section \ref{sec:toy}: we compare the mean of the estimators from noisy images for the case when the true image is a real galaxy, relative to the case when the true image is a best fit of our model to the real galaxy without added noise.
This time, we use more a realistic fitted galaxy model, consisting of two components: bulge and disc. 
Bulge component is a de Vaucouleurs profile with S\'{e}rsic index $n=4$, disc is an Exponential profile with S\'{e}rsic index $n=1$, ratio of the half light radii of the components was set to $r_B/r_D=1$.
This model was previously used in \citet{im3shape} and K12.

This time, however, we can not afford to repeat this procedure for all 26113 galaxies available in the COSMOS sample, due to computational reasons;
constrainig the multiplicative bias to $\sigma_m < 0.004$ for a single galaxy image with $S/N=20$ requires of order 4 million noise realisations.
Instead, we group those galaxies in bins of size, redshift, morphological classification, and also model bias.
For each galaxy in a bin we use a number of ring tests, with different shears. 

We create a set of 8 shears by using all pairs of $g_1 , g_2 \in \{-0.1, 0.0, 0.1\}$, except for $g_1=g_2=0.0$.
The reason for missing out this middle point is that it brings very little statistical power for constraining the multiplicative bias, whereas additive bias is already constrained quite well.
For each shear we create a ring test with 8 equally spaced angles. 

Our simulation data set consisted of almost 20 million galaxy images; half of them were real galaxy images, the other half their best S\'{e}rsic representations.
The number of noise realisations per COSMOS galaxy image was variable and dependent on the number of galaxies in bins of redshift and morphological classification.
There are very few galaxies with high redshift or low Hubble Sequence index available in the COSMOS sample, so the number of noise realisations for each of these galaxies had to be larger than for others to reach the desired statistical uncertainty for these bins.

The image pixel size was 0.27 arcsec, and we fitted to postage stamps of size $39 \times 39$ pixels.
We use a Moffat profile for the target reconvolving PSF, with FWHM of $0.7695$ arcsec$=2.85$ pixels, and $\beta=3$. 
The ellipticity of the PSF was $g_1^{\mathrm{PSF}}=g_2^{\mathrm{PSF}}=0.05$.
When we added noise to the simulated images, the signal to noise ratio was again $S/N=20$.

We measure the ratio of the FWHM of the convolved galaxy (without noise) to the FWHM of the PSF ($\size$).
FWHM of the PSF-convolved galaxy is measured numerically from an image of the best fitting bulge plus disc model drawed on a fine grid, with both galaxy and PSF ellipticities set to zero.
For small galaxies the shear is strongly underestimated
due to noise bias. 
Multiplicative bias for galaxies in size bin of $\size \in (1.2,1.3)$ is $m_i \sim -0.2$, whereas model bias is still on sub-percent level.
Therefore we remove all galaxies with $\size < 1.3$ for the purpose of this work.

If we perform a reconvolution and shearing operation on a galaxy image containing noise, the final image will contain correlated noise which may align with the shear direction.
To assure that this effect is not dominant in our results, we only consider COSMOS galaxies for which $S/N>200$.

We compared the bias obtained from real galaxy images (including the interaction terms) and best-fitting images (excluding interaction terms) for the whole sample available in \textsc{\textsc{GalSim}}.
Table \ref{tab:total_bias} shows the differences between those. 
The results indicate that for the entire selected sample the difference is positive and on a very small level of $m \approx 0.001$ , and only $1\sigma$ significant.

The mean model bias measured by \textsc{Im3shape} with a bulge + disc model was of order $m = 0.005$ and $c = 0.0003$. 
Additionally, we measured the model bias for each individual galaxy in the sample and obtained standard deviation on multiplicative bias $\std{m} = 0.02$. 
Model bias can vary significantly, but happens to average out to a rather small value.

\begin{figure*}
\epsfig{file=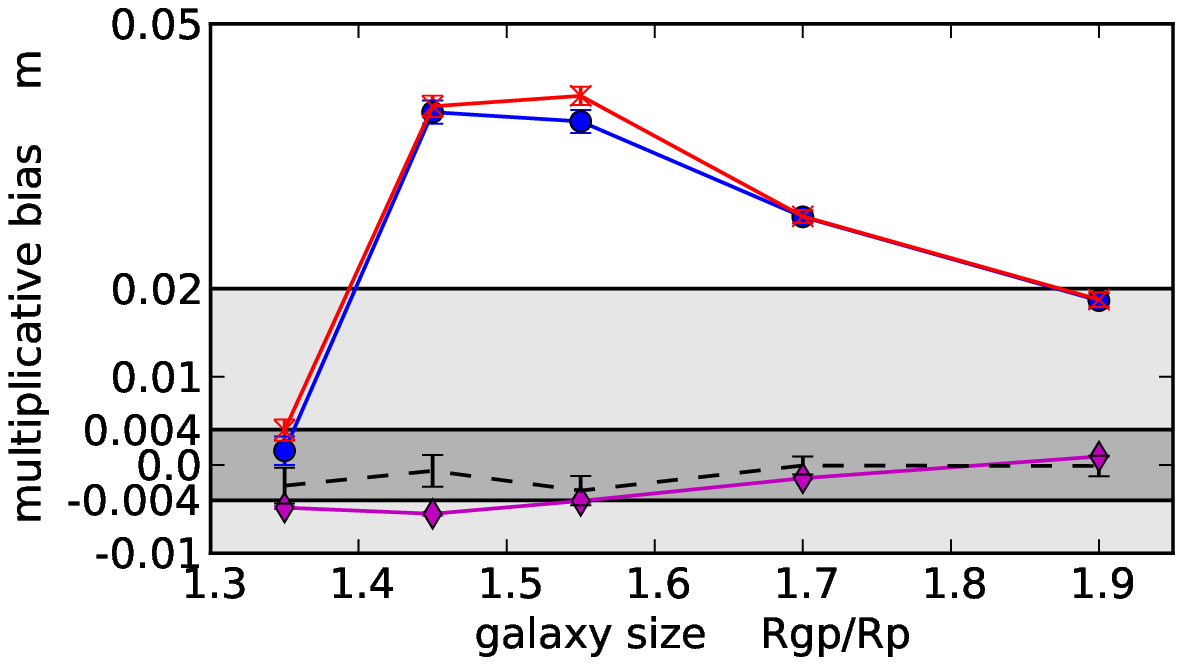             		 ,width=8.5cm}
\epsfig{file=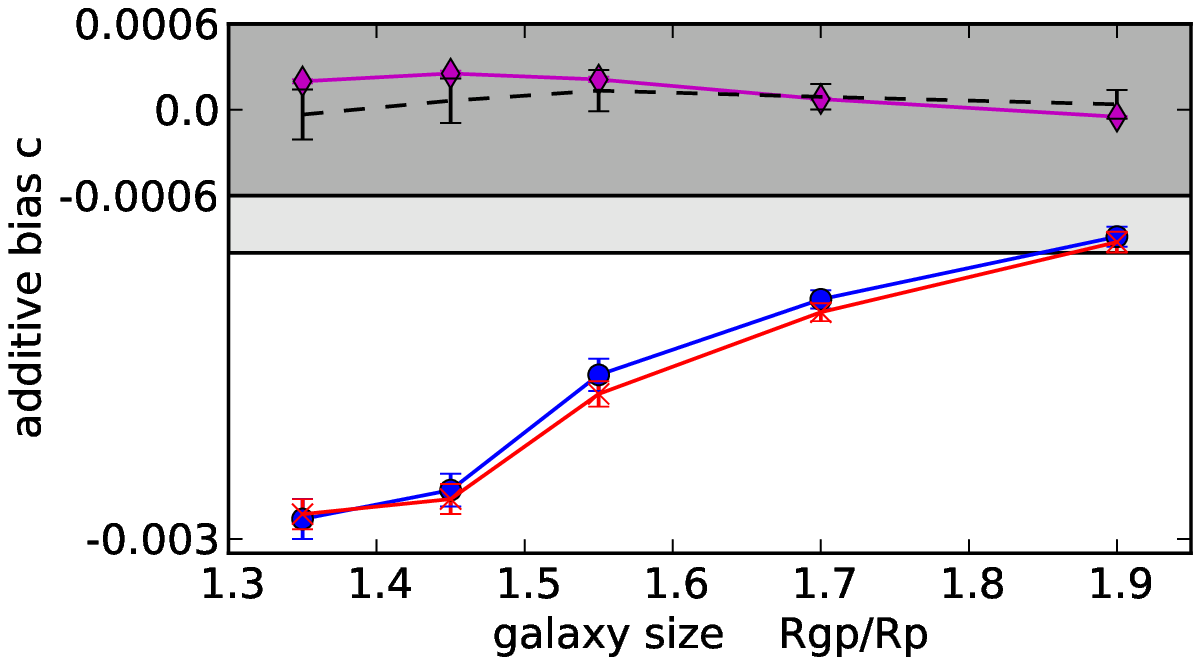    		 ,width=8.5cm}
\epsfig{file=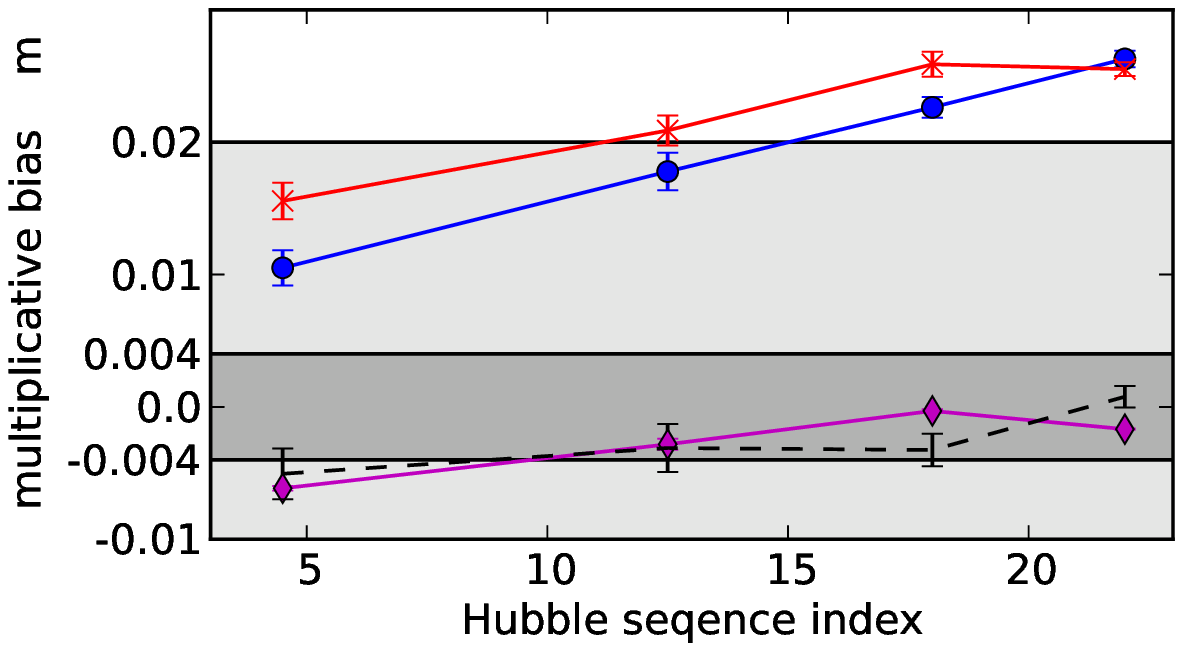             		 ,width=8.5cm}
\epsfig{file=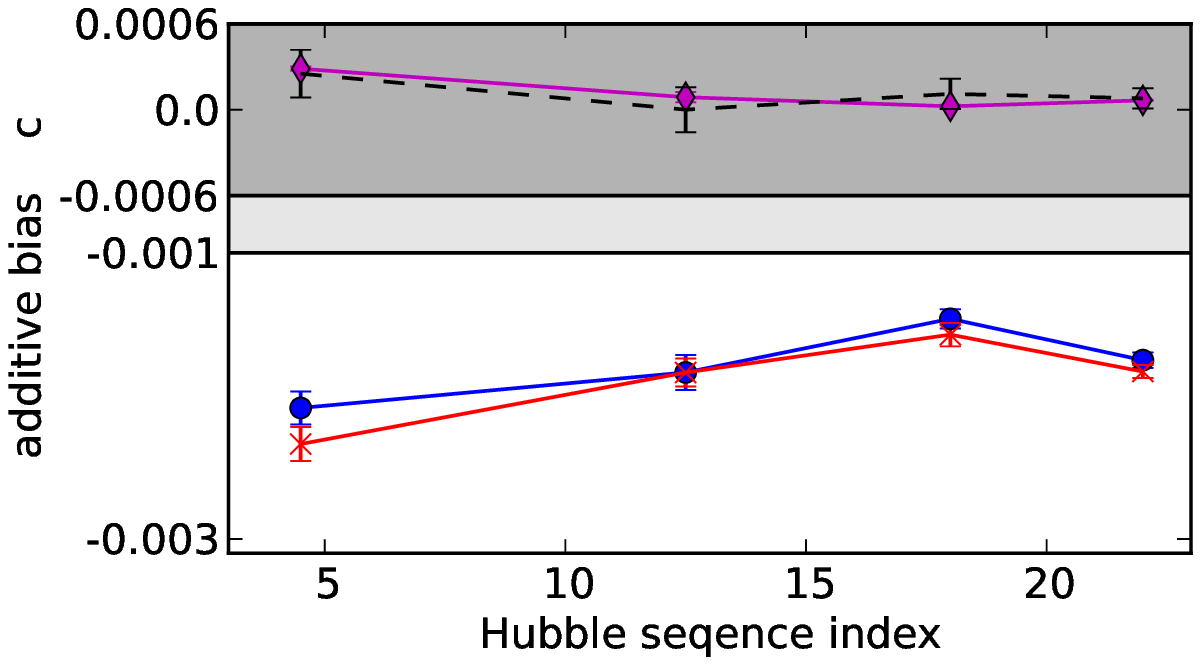    		 ,width=8.5cm}
\epsfig{file=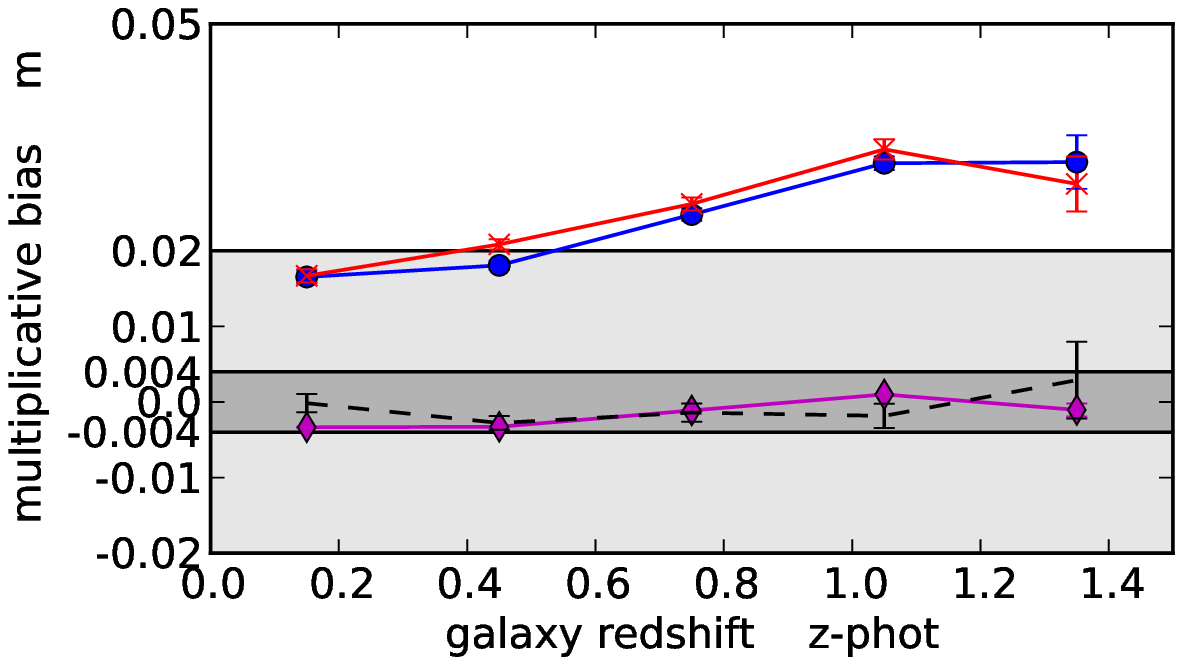         	 ,width=8.5cm}
\epsfig{file=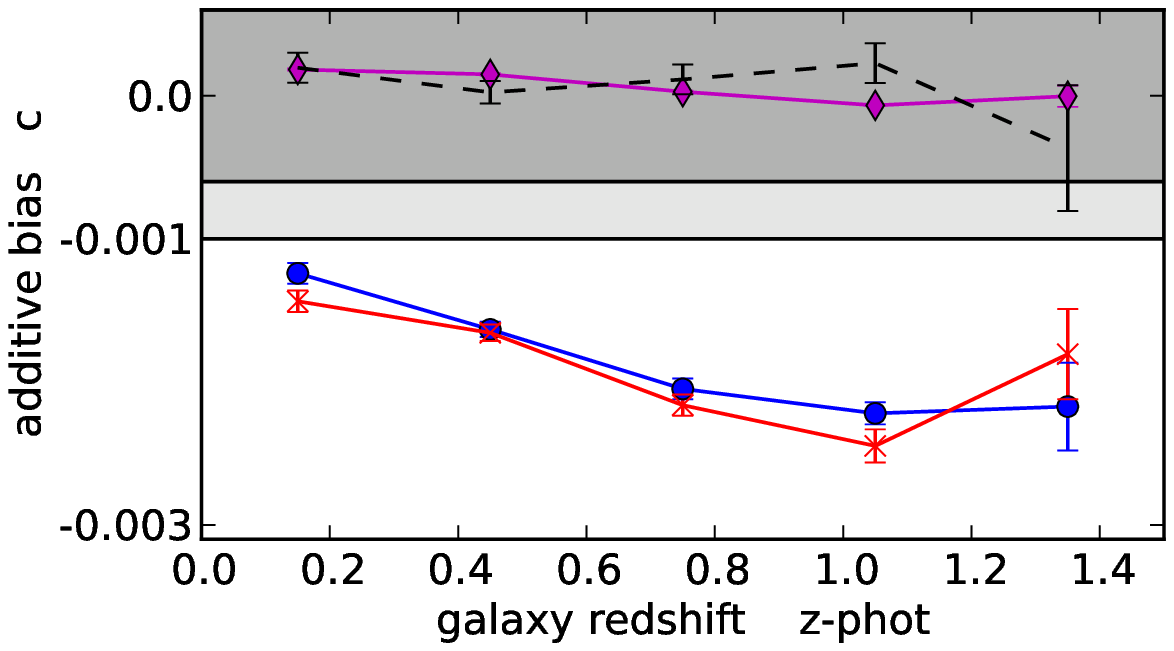	 ,width=8.5cm}
\epsfig{file=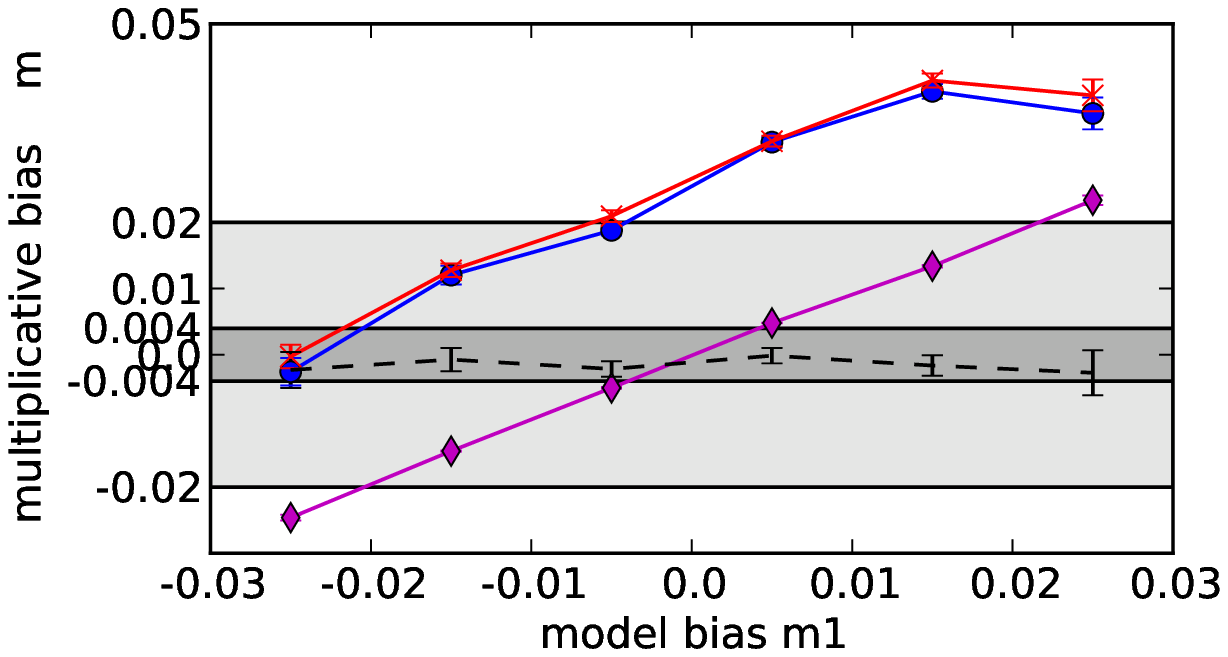 ,width=8.5cm}
\epsfig{file=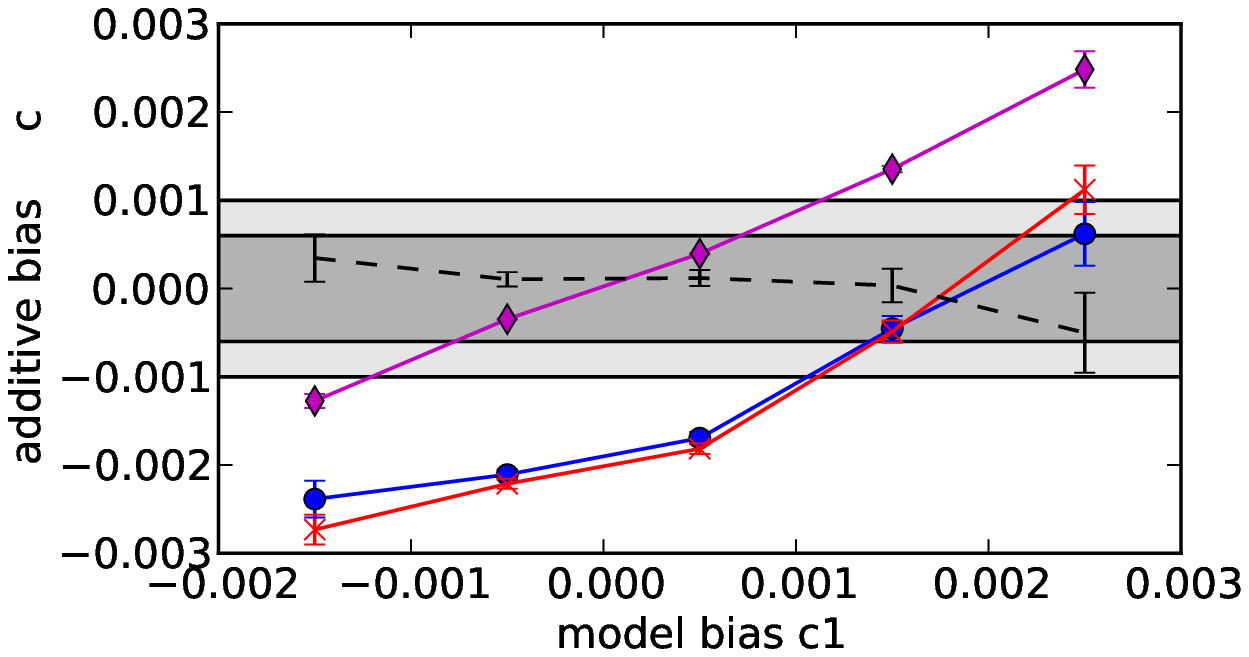 ,width=8.5cm}
\epsfig{file=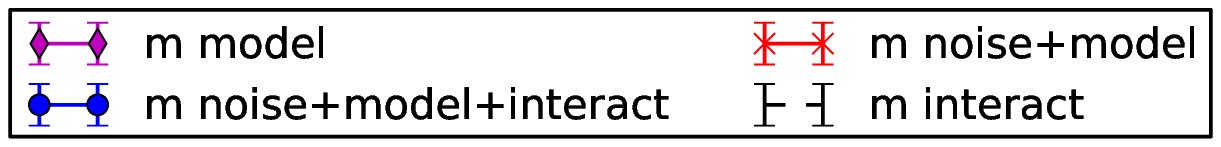 		 ,width=8cm}
\epsfig{file=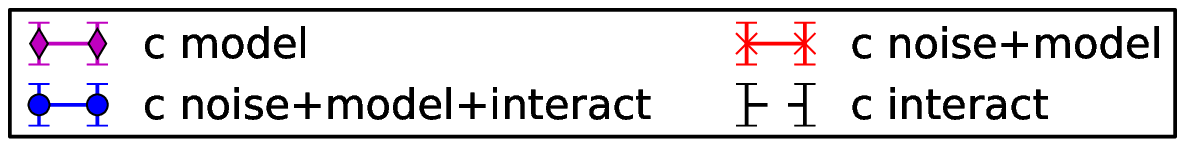 		 ,width=8cm}
\caption{
Model bias, noise bias and their interaction as a function of various galaxy properties.
Left and right panels show multiplicative and additive biases, respectively.
The biases shown here are the average of two components: $m=(m_1+m_2)/2$, ditto for additive bias.
The series on these plots presents: 
(i)  the model bias, obtained from noiseless real galaxy images from the COSMOS catalogue, available in \textsc{GalSim} (magenta diamonds),
(ii) model bias + noise bias, calculated from noise realisations, when the true galaxy was a bulge + disc S\'{e}rsic
 model (red crosses). The parameters of that model were that of a best-fit to the noiseless real galaxy,
(iii) model bias + noise bias + their interaction, calculated from noise realisations, when the true galaxy was the real COSMOS galaxy image (blue circles).
The upper left panel shows the biases as a function of galaxy size, measured with respect to the PSF size (FWHM of the convolved galaxy divided by the FWHM of the PSF).
The upper right panel shows the biases as a function of the Hubble Sequence Index, where the index corresponds to types:
$1\dots8$ Ell-S0, $9\dots15$ Sa-Sc, $16\dots19$ Sd-Sdm, $\geq 20$ starburst.
The lower left panel shows the biases as a function of galaxy photometric redshift.
The lower right panel shows the biases as a function of the value of the galaxy model bias.
\label{fig:eval}
}
\end{figure*}

In Fig. \ref{fig:eval} we show the model bias (magenta diamonds), model bias + noise bias (red crosses), and model bias + noise bias + their interaction (blue circles), for galaxies binned by different properties.
We also show directly the strength of the interaction terms (black dashed).
The bias obtained from the noisy realisations, where the noiseless image was a real galaxy image from the COSMOS sample (blue circles)
and its best S\'{e}rsic representation (red crosses). 

The upper left panel shows those biases as a function of the true size of the real galaxy, measured as $\size$. 
For more discussion and investigation of the model bias results alone  see
Voigt et al. (2013, in prep.). 
Estimates from galaxies with $\size > 1.3$ seem to be biased positive in the presence of noise.
The noise bias decreases as the size of the galaxy increases.
This dependence is similar to the one presented in K12.
For a galaxy with $\size=1.6$, in K12 the noise bias was $m=0.039$ for disc galaxies, $m=-0.013$ for a bulge galaxies and $m=0.02$ for galaxies with bulge-to-disc ratio of 1.
Here, galaxies with this size are biased on the level of $m=0.035$.

The difference between noise bias obtained from the real galaxies and their best-fitting S\'{e}rsic models is shown using the black dashed lines.
This difference shows directly the strength of the noise and model bias interaction terms.
The difference is very small, consistent with zero to the accuracy limit imposed by the finite number of noise realisations we simulated. 
This is the case for all the size bins.
This means that these terms are either small enough not to make a difference, or their contribution averages out, which should therefore also make no difference to shear statistics, to first order.

The upper middle panels show the multiplicative and additive bias as a function of Hubble Sequence index. 
The index corresponds to the following types: $1\dots8$ Ell-S0, $9\dots15$ Sa-Sc, $16\dots19$ Sd-Sdm, $\geq 20$ starburst from \citep{BruzualCharlot2003}.
The majority of galaxies in our COSMOS sample have Hubble index $>9$. 
For those galaxies, we can not see a significant difference between noise bias from real galaxy images and their best S\'{e}rsic representations.
For the elliptical galaxies, however, see a $1-2\sigma$ deviation from zero. 
This difference is still small within the requirements for upcoming galaxy surveys, although they would dominate the error budget.

The noise and model bias interaction does not significantly deviate from zero as a function of galaxy redshift, for either additive or multiplicative bias.

The model bias should be dependent on the residuals of the best fit image to the real galaxy image, and so should the noise and model bias interaction.
We wanted to investigate if there is a dependence between those two effects.
Therefore the bottom panels show the biases as a function of model bias. 
It seems that the noise and model interaction bias terms do not depend strongly on the model bias, which may indicate that their contribution averages out to a very small value, for this galaxy sample.

\section{Discussion and conclusions}
\label{sec:summary}
 
We investigated the problem of shape estimation using model fitting methods, studying in detail the impact of using a model which does not represent the galaxy morphology completely, with noise included.
For noiseless galaxies, \citet{bernstein2010,Voigt2011} introduced and investigated the model bias, which will depend on the complexity of the fitted model.
Noise bias, studied in R12 and K12 was derived for the case when the true noiseless galaxy \emph{can} be represented perfectly by our fitted model - there was no bias in the absence of noise.
This bias depends on the signal to noise of the image, as well as galaxy size and other properties.

In this work we generalised the noise bias derivations to the case when an imperfect model is used.
The interaction between the noise and model bias depends on the residual of the best fit model and the real galaxy image.
We isolated the terms dependent on this residual, by comparing the noise bias arising when the true underlying image is the \emph{real} galaxy (residual present), and when it is a \emph{best-fitted} S\'{e}rsic image (residuals absent).
We call this difference a \emph{noise and model bias interaction}. 
Thus, model bias, noise bias and their interaction make a complete picture of biases we can expect for fitting parametric models to isolated real galaxy images.

A simple toy model was shown to demonstrate the potential influence of these terms. 
In our simple toy model, we encountered model biases on a level of $20\%$, noise biases of order $10\%$, and the interaction terms of order $1-2\%$. 
For the toy model, the interaction terms are small compared to model and noise biases, but significant enough that we can detect it using a simulation with finite number of noise realisations.

We then investigated the shear bias induced by model fitting for the real galaxy images available in a COSMOS sample \citep{shera}.
The mean model bias measured by \textsc{Im3shape} with a bulge + disc model was small; of order $m = 0.005$ and $c=0.0003$.
The standard deviation of model bias for individual galaxies is higher; on the level of $\std{m} = 0.02$.
This indicates that the model bias can be quite variable, but it averages to a very small value.

For the signal to noise level of $S/N=20$, the maximum noise bias we obtained was on the level of $m \sim 0.04$ and $c \sim -0.003$, for galaxies falling into a size bin of $\size \sim (1.4,1.6)$.
As a function of galaxy size, the noise bias dependence is very compatible with K12; it decreases as the galaxy size increases. 
The magnitude of this bias corresponds to a galaxy sample with approximately 2/3 disc-like and 1/3 bulge-like galaxies.

Noise bias as function of other galaxy properties, such as redshift and morphological classification, is also variable.
However this is almost certainly due to the correlation of the property with galaxy size, which influences the noise bias the most.

The noise and model bias interaction terms prove to be very small, usually almost consistent with zero to the accuracy we simulated, which varied from $\sigma_{m}\sim 0.001$ to $\sigma_{m} \sim 0.006$ across bins in different parameters. 
The only deviation from zero was found for galaxies with Hubble Sequence Index $<19$, which corresponds to ellipticals.
The noise and model bias interaction terms were estimated as $m=0.004 \pm 0.002$, which is a $2\sigma$ deviation from zero and on the borderline of shortly  upcoming survey requirements.

The most important shear bias dependence is on galaxy redshift, as it impacts the shear tomography the most. 
We split our galaxy sample into bins of redshift and measured the model and noise biases for each bin. 
Noise and model bias interaction does not seem to deviate more than $1-2\sigma$ from zero for each bin. 

This indicates that the bulge and disc S\'{e}rsic model should be a good basis for noise bias calibrations for current surveys.
Furthermore, the model bias and noise bias calibrations can be performed in two separate steps.
Using S\'{e}rsic profile parameters to evaluate the noise bias can reduce the simulation volume significantly, by using a procedure described in K12. 
Similarly, if noise bias equations can be evaluated analytically or semi-analytically with sufficient precision, the fact that the noise and model bias interaction terms can be neglected to the accuracy of roughly $\sigma_{m} \sim 0.003$, means that the noise bias prediction task can be simplified even further.
Also, if a galaxy model used proves to be robust enough and introduce negligible model bias, simulations based on S\'{e}rsic models should prove sufficient for the shear calibration.

For the results presented in this paper we used a S\'{e}rsic, co-elliptical, co-centric bulge + disc model, with fixed radii ratio of the two components.
This model choice is well motivated, but arbitrary to some extent. 
Selecting a good model for a galaxy is not a trivial task.
Increasing the model complexity by adding more parameters (for example allowing for different, variable ellipticities for bulge and disc components) can result in lower model biases \citep{Voigt2011}.

For upcoming and far future surveys, the noise and model bias interaction terms might become important. 
Minimizing this contribution may involve finding a better galaxy model (simultaneously decreasing the model bias), 
and/or using representative calibration images and reconvolution technique. 
When a good quality, representative image sample is available, or model bias calibration is necessary, it will always be more accurate to use the images themselves.

\section*{Acknowledgements}
TK, SB, MH, BR and JZ acknowledge support from the European Research Council in the form of a Starting Grant with number 240672. 
We thank Alexandre Refregier and Adam Amara for helpful discussions and suggestions.
We thank Mike Jarvis, Rachel Mandelbaum, Gary Bernstein and other developers of \textsc{GalSim} software for excellent support. 
The authors acknowledge the use of the UCL Legion High Performance Computing Facility (Legion@UCL), and associated support services, in the completion of this work.

\appendix

\section{Detailed derivation of the noise and model bias interaction terms}
\label{sec:appendix1}

Let's define the following set of variables

\begin{align}
g_p 	     \quad & \text{- true noiseless image } \\
\vec{a}^t    \quad & \text{- vector of best fitting parameters for $g_p$} \\
f_p(\vec{a}) \quad & \text{- model image with parameters $\vec{a}$ at pixel $p$} \\    
n_p	     \quad & \text{- noise at pixel $p$ } \\
\sigma_n     \quad & \text{- noise standard deviation } \\
\rho=\frac{f_0}{\sigma_n}     \quad & \text{- signal to noise, assume total flux $f_0=1$ } 
\end{align}
Log - likelihood of image data given a set of parameters is
\begin{align}
\label{chi2}
-2 \log{{\cal L}} = \chi^2(\vec{a}) = \frac{1}{\sigma^2} \sum_p \left[ g_p + n_p - f_p(\vec{a}) \right]^2  
\end{align}
Maximum likelihood point $\hat{\vec{a}}$ is defined as
\begin{align}
\hat{\vec{a}} = \argmin_{\vec{a}} \chi^2(\vec{a}) \quad \Leftrightarrow \quad \frac{\partial \chi^2(\vec{a}) }{ \partial a_k } = 0 \quad \forall \quad k 
\end{align}
where
\begin{align}
\label{eq:dchi2}
\frac{\partial \chi^2(\vec{a}) }{ \partial a_k } 
=
\frac{2}{\sigma^2} \sum_p \left[ \left[ g_p + n_p - f_p(\vec{a}) \right] (-1) \frac{\partial f_p(\vec{a}) }{ \partial a_k }  \right]
\end{align}
An expansion of $\hat{\vec{a}}$ around the true parameters $\vec{a}^t$ gives
\begin{align}
\hat{a}_k = a_k^t + \underbrace{\alpha a_k^{(1)} + \alpha^2 a_k^{(2)}}_{\delta a_k}
\label{expansion_a} 
\end{align}
Expanding $f_p(\hat{\vec{a}})$ about $\vec{a}^t$ gives
\begin{align}
f_p(\hat{\vec{a}}) 
= 
  f_p(\vec{a}^t) 
+ \sum_i \delta a_i \frac{\partial f_p(\vec{a}^t) }{ \partial a_i } 
+ \frac{1}{2} \sum_{ij} \delta a_i \delta a_j \frac{\partial^2 f_p(\vec{a}^t) }{ \partial a_i \partial a_j }
\label{expansion_f}  
\end{align}
And finally, expanding $\frac{\partial f_p(\hat{\vec{a}}) }{ \partial a_i }$ gives
\begin{align}
\label{expansion_df}
\frac{\partial f_p(\hat{\vec{a}}) }{ \partial a_k } 
=                                                     
  \frac{\partial f_p(\vec{a}^t) }{ \partial a_k }
+ \sum_i \delta a_i \frac{\partial^2 f_p(\vec{a}^t) }{ \partial a_k a_i }   
+ \frac{1}{2} \sum_{ij} \delta a_i \delta a_j \frac{\partial^3 f_p(\vec{a}^t) }{ \partial a_k a_i a_j}
\end{align}
Now we substitute (\ref{expansion_a}), (\ref{expansion_f}), (\ref{expansion_df}) into (\ref{eq:dchi2}), ignore terms $O(\alpha>2)$, and use the residual $r_p:=g_p - f_p(\vec{a}^t)$.
For convenience let's use $ \frac{\partial f_p(\vec{a}^t)}{\partial a_k} = \frac{\partial f^t_p}{\partial a_k}$, and then we can write
\begin{align}
\label{chi2_3}
\frac{\partial \chi^2(\vec{a}) }{ \partial a_k } 
= \nonumber \\
\frac{-2}{\sigma^2} 
\sum_p  \Bigg( 
\alpha &\Bigg[ 
  {n_p \frac{\partial f^t_p }{ \partial a_k }} 
- {\sum_i a_i^{(1)} \frac{\partial f^t_p }{ \partial a_i } \frac{\partial f^t_p }{ \partial a_k }}
+ {r_p \sum_{i} a_i^{(1)} \frac{\partial^2 f^t_p }{ \partial a_i \partial a_k  } }
\Bigg]  \nonumber \\
+ \alpha^2 
&\Bigg[ 
  {n_p \sum_i a_i^{(1)} \frac{\partial^2 f^t_p }{ \partial a_i \partial a_k }} 
- { \sum_{ij} a_i^{(1)} a_j^{(1)} \frac{\partial f^t_p }{ \partial a_i } \frac{\partial^2 f^t_p }{ \partial a_j \partial a_k }}
\nonumber \\ 
& -{\sum_i a_i^{(2)} \frac{\partial f^t_p }{ \partial a_i } \frac{\partial f^t_p }{ \partial a_k }} 
  -{\frac{1}{2} \sum_{ij} a_i^{(1)} a_j^{(1)} \frac{\partial^2 f^t_p }{ \partial a_i \partial a_j } \frac{\partial f^t_p }{ \partial a_k }}
\nonumber \\ 
& +{r_p \sum_i a_i^{(2)} \frac{\partial^2 f^t_p }{ \partial a_i \partial a_k } } 
 + {\frac{1}{2} r_p \sum_{ij} a_i^{(1)} a_j^{(1)} \frac{\partial^3 f^t_p }{ \partial a_i \partial a_j \partial a_k } }
\Bigg] 
\Bigg)
\end{align}
Using definitions in Eqn. \ref{eqn:D_def}
and summation over repeated indices, we can collect the first order terms
\begin{align}
O(\alpha) = 0 \quad  \Leftrightarrow \quad  D^{(1)}_{kp} n_p -  F_{ik}  a^{(1)}_i + r_p D^{(2)}_{ikp} a^{(1)}_i = 0\quad \forall \quad a_k
\end{align}
Solving for first order terms
\begin{align}
\label{a1}
a^{(1)}_i &= (F_{ik} - r_p D^{(2)}_{ikp})^{-1} D^{(1)}_{kp}n_p \\
\langle a^{(1)}_i \rangle &= 0
\end{align}
As expected from the result of \citep{NoiseBias1}, the first order terms average to zero.
For convenience, we can use a Fisher matrix modified by the residual $r_p$
\begin{align}
\tilde F_{ik} := (F_{ik}  - r_p D^{(2)}_{ikp}) 
\end{align}
Covariance between two parameters is 
\begin{align}
\langle a^{(1)}_i a^{(1)}_j \rangle &=  \tilde F_{ij} ^{-1} D^{(1)}_{jp} \underbrace{\langle n_p n_p \rangle}_{\sigma^2} D^{(1)}_{jp}  \tilde F_{ij} ^{-1} \\
\label{cov_a1}
&= \sigma^2 \tilde F_{ij}^{-1} F_{ij} \tilde F_{ij}^{-1}
\end{align}
Covariance between a parameter and a noise pixel, using (\ref{a1}), is
\begin{align}
\label{cov_a1np}
 \langle a^{(1)}_i n_m \rangle = \langle \tilde F_{ij}^{-1}  D^{(1)}_{jp} n_p n_m \rangle = \tilde F_{ij}^{-1} D^{(1)}_{jp} \delta(m=p) \sigma_n^2   
\end{align}
Solving for second order terms in (\ref{chi2_3}) gives
\begin{align}
 O(\alpha^2) = 0 \quad  \Leftrightarrow \quad
&   a^{(1)}_i D^{(2)}_{ikp} n_p \nonumber \\
& + a^{(1)}_i a^{(1)}_j \Bigg(  \frac{1}{2} r_p D^{(3)}_{ijkp}  - D^{(1)}_{jp} D^{(2)}_{ikp} -  \frac{1}{2}D^{(1)}_{kp}D^{(2)}_{ijp} \Bigg)\nonumber \\  
& - a^{(2)}_i  \tilde F_{ik} \qquad = 0
  \quad \forall \quad a_k
 \end{align}
Averaging second order terms, rearranging and substituting (\ref{cov_a1np}) and (\ref{cov_a1}) we arrive at Eqn. \ref{eqn:nmb_final}.
This is our main result, showing how the residual between the best fit and the noiseless real galaxy image is modifying the noise bias equations.
It introduces the third derivative of the model function scaled by the residual  $D^{(3)}_{ljkp} r_p$.
It also modifies the Fisher matrix, and may prevent the cancellation of two terms in $D^{(1)}_{jp} D^{(2)}_{lkp}$.
If there is no model bias, and the residual $r_p = 0$, and the expression reduces to the result from \citep{NoiseBias1} shown in Eqn. \ref{eqn:nb_final}.
\label{lastpage}

\end{document}